\documentclass[preprint,12pt]{elsarticle}

\usepackage{nomencl}
\RequirePackage{ifthen}
 \renewcommand{\nomgroup}[1]{%
 \ifthenelse{\equal{#1}{R}}{\item[\textbf{Roman symbols}]}{%
 \ifthenelse{\equal{#1}{G}}{\item[\textbf{Greek symbols}]}{%
 \ifthenelse{\equal{#1}{A}}{\item[\textbf{Abbreviations}]}{%
 \ifthenelse{\equal{#1}{S}}{\item[\textbf{Subscript}]}{%
 \ifthenelse{\equal{#1}{T}}{\item[\textbf{Superscripts}]}{}}}}}
}\makenomenclature
\nomenclature[r]{$B$}{Skempton coefficient ($-$)}%
\nomenclature[r]{$K_b$}{drained bulk modulus ($Pa$)}%
\nomenclature[r]{$K_p$}{drained pore modulus ($Pa$)}%
\nomenclature[r]{$K_s$}{unjacketed bulk modulus ($Pa$)}%
\nomenclature[r]{$K_\phi$}{unjacketed pore modulus ($Pa$)}%
\nomenclature[r]{$K_f$}{fluid modulus ($Pa$)}%
\nomenclature[r]{$V_p$}{pore volume ($m^3$)}%
\nomenclature[r]{$V_p^i$}{initial pore volume ($m^3$)}%
\nomenclature[r]{$V_b$}{bulk volume ($m^3$)}%
\nomenclature[r]{$V_b^i$}{initial bulk volume ($m^3$)}%
\nomenclature[r]{$p_p$}{pore pressure ($Pa$)}%
\nomenclature[r]{$p_c$}{confining pressure ($Pa$)}%
\nomenclature[r]{$p_e$}{$p_c-p_p$ effective pressure ($Pa$)}%
\nomenclature[r]{$C_b$}{drained bulk compressibility ($1/Pa$)}%
\nomenclature[r]{$C_p$}{drained pore compressibility ($1/Pa$)}%
\nomenclature[r]{$C_s$}{unjacketed bulk compressibility ($1/Pa$)}%
\nomenclature[r]{$C_\phi$}{unjacketed pore compressibility ($1/Pa$)}%
\nomenclature[r]{$C_f$}{fluid compressibility ($1/Pa$)}%
\nomenclature[r]{$C_m$}{solid matrix compressibility ($1/Pa$)}%
\nomenclature[r]{$n_{eff}$}{refractive index of fibre core material ($-$)}%
\nomenclature[r]{$n$}{refractive index of extrinsic fabry perot cavity ($-$)}%
\nomenclature[r]{$L$}{length of extrinsic fabry perot cavity ($m$)}%
\nomenclature[r]{$L_s$}{distance between fusion splices ($m$)}%
\nomenclature[r]{$\Delta L_T$}{length change due to temperature ($m$)}%
\nomenclature[r]{$\Delta L_p$}{length change due to pressure ($m$)}%
\nomenclature[r]{$R$}{reflection coefficient ($-$)}%
\nomenclature[r]{$r_{i}$}{inner radius of glass capillary ($m$)}%
\nomenclature[r]{$r_{o}$}{outer radius of glass capillary ($m$)}%
\nomenclature[r]{$E$}{Young's modulus ($Pa$)}%
\nomenclature[r]{$p$}{pressure ($Pa$)}%
\nomenclature[r]{$T$}{temperature ($K$)}%
\nomenclature[r]{$a_{ii}$}{auxilliary function with i=[1,2] ($-$)}%
\nomenclature[r]{$I_0$}{intensity of incident light ($W/m^2$)}%
\nomenclature[r]{$I_R$}{intensity of reflected light ($W/m^2$)}%
\nomenclature[r]{$m_f$}{fluid mass ($kg$)}%

\nomenclature[g]{$\alpha$}{Biot coefficient ($-$)}%
\nomenclature[g]{$\beta_C$}{coefficient of thermal expansion (capillary) ($1/K$)}%
\nomenclature[g]{$\beta_F$}{coefficient of thermal expansion (fibre) ($1/K$)}%
\nomenclature[g]{$\epsilon{}_v$}{volumetric strain ($-$)}%
\nomenclature[g]{$\epsilon{}_a$}{axial strain ($-$)}%
\nomenclature[g]{$\epsilon{}_c$}{circumferential strain ($-$)}%
\nomenclature[g]{$\phi{}$}{porosity ($-$)}%
\nomenclature[g]{$\phi_L$}{Lagrange's porosity ($-$)}%
\nomenclature[g]{$\phi_E$}{Euler's porosity ($-$)}%
\nomenclature[g]{$\phi{}^i$}{initial porosity ($-$)}%
\nomenclature[g]{$\lambda$}{wavelength ($nm$)}
\nomenclature[g]{$\lambda_{FSR}$}{free spectral range ($nm$)}
\nomenclature[g]{$\Lambda$}{period of Fibre Bragg Grating ($nm$)}
\nomenclature[g]{$\lambda_B$}{Bragg wavelength ($nm$)}
\nomenclature[g]{$\varphi_C$}{phase shift between reflected signals ($-$)}
\nomenclature[g]{$\mu$}{Poisson's ratio ($-$)}%

\nomenclature[a]{$SM$}{single mode}%
\nomenclature[a]{$FOPS$}{fibre optic pressure sensor}%
\nomenclature[a]{$EFPI$}{Extrinsic Fabry-Perot Interferometer}%
\nomenclature[a]{$FBG$}{Fibre Bragg Grating}%
\nomenclature[a]{$CTE$}{coefficient of thermal expansion}%
\nomenclature[a]{$FFT$}{Fast Fourier Transformation}%
\nomenclature[a]{$FRFT$}{Fractional Fourier Transformation}%
\nomenclature[a]{$hp$}{high pressure}%
\nomenclature[a]{$hT$}{high temperature}%
\nomenclature[a]{$FEP$}{fluorinated ethylene propylene}
\nomenclature[a]{$OSA$}{optical spectrum analyser}
\nomenclature[a]{$BBS$}{broad band light source}
\nomenclature[a]{$HTMC$}{hydraulic, thermal, mechanical and chemical}

\usepackage{graphics}

\usepackage{amssymb}

\usepackage{natbib}

\usepackage{lineno}

\usepackage{booktabs}
\usepackage{subfigure}
\usepackage{color}
\usepackage{amsmath}
\usepackage[ansinew]{inputenc}

\journal{IJRMMS}
\begin{document}
NOTICE: This is the author’s version of a work that was
accepted for publication in International Journal of Rock Mechanics and Mining Sciences. Changes resulting
from the publishing process, such as peer review, editing,
corrections, structural formatting, and other quality control
mechanisms, may not be reflected in this document. Changes
may have been made to this work since it was submitted for
publication. A definitive version was subsequently published in International Journal of Rock Mechanics and Mining Sciences Volume 55, October 2012, Pages 55–62, http://dx.doi.org/10.1016/j.ijrmms.2012.06.011 .
\clearpage
\begin{frontmatter}

\title{A fibre optic sensor for
the in situ determination of rock physical properties\tnoteref{technote}}
\tnotetext[technote]{Technical Note}
\author[gfz]{Thomas Reinsch\corref{cor1}}
\ead{Thomas.Reinsch@gfz-potsdam.de}
\author[gfz]{Guido Bl\"{o}cher}
\author[gfz]{Harald Milsch}
\author[hot]{Kort Bremer}
\author[ui]{Elfed Lewis}
\author[ui]{Gabriel Leen}
\author[uw]{Steffen Lochmann}

\cortext[cor1]{Corresponding author}
\address[gfz]{Helmholtz Centre Potsdam, GFZ German Research Centre for Geosciences, Telegrafenberg, 14473 Potsdam, Germany}
\address[hot]{HOT-Hanover Center for Optical Technologies, Leibniz Universit\"{a}t Hannover, Hannover, Germany }
\address[ui]{Optical Fibre Sensor Research Centre (OFSRC), University of Limerick, Limerick, Ireland}
\address[uw]{Dept. of Electrical Eng. and Computer Science, Hochschule Wismar, Wismar, Germany}


\end{frontmatter}

\section{Introduction}
\label{intro}
The hydraulic, thermal, mechanical and chemical (HTMC) characterisation of rocks is essential for the understanding of subsurface processes like earthquake mechanics or processes that result from the production or injection of fluids into subsurface reservoirs. Injecting into, or producing fluids from a reservoir has an effect on pressure, temperature and chemical equilibrium of the reservoir rocks. From the theory of poroelasticity it is known that when a porous medium is subjected to different load conditions, this results in a matrix deformation, which leads to volumetric changes of the pore space and the surrounding grains. Furthermore, the pore pressure will change and the fluid starts to flow from regions of high to low pressure.

In order to perform the necessary laboratory experiments to characterize the HTMC properties of rock specimens, several parameters have to be measured at different locations along the sample. Mechanical deformation or hydrostatic pressure conditions are measured outside the sample, whereas e.g.\ pore fluid properties are measured within the fluid system, connected to the pore space of the specimen under investigation. In undrained hydrostatic compression tests, for example, where a jacketed rock specimen is subjected to varying load conditions, the pore pressure change is a measure of the stiffness of the rock matrix. Pore pressure measurements, however, are often biased due to the experimental set-up where large tubing volumes, compared to the volume of the pore space, are needed to connect the pore space with conventional pressure gauges. This additional volume in the exterior tubing can significantly influence measurements of undrained pore pressure \cite{Green1986}, because the external saturant volume acts as a pore fluid reservoir and violates the undrained condition $dm_f=0$. In addition, the compressibility and the thermal expansion of the exterior tubing and of the dead volume in the fluid filling system can influence the undrained pore pressure measurement. This influence can be corrected following the procedure outlined in \citet{Ghabezloo2010}. Correction methods were previously presented by \citet{Wissa1969}. Furthermore, Wissa et al. \cite{Wissa1969} suggested to reduce the total volume of pore pressure lines in the triaxial cell base on less than 3\% of the pore volume in the test specimen.

To overcome the limitation of a large void space, a novel fibre optic pressure and temperature gauge has been embedded directly within the sample and pore pressure measurements have been performed. Although fibre optic techniques have been used to measure mechanical deformation of a sample within a triaxial cell \citep[e.g.][]{Schmidt-Hattenberger2003}, to our knowledge, pore pressure measurements have been performed for the first time in this study.

Within this contribution, the experimental procedure will be described and the measured rock physical parameters will be outlined. The accuracy and applicability of the sensor will be discussed in detail. The measured poroelastic parameters will be discussed in a subsequent publication.

\section{Experimental Set-Up - Sensing}
\label{sec:exp}
\subsection{Sensing Principle}
\begin{figure}%
\centering
\includegraphics[width=0.6\textwidth]{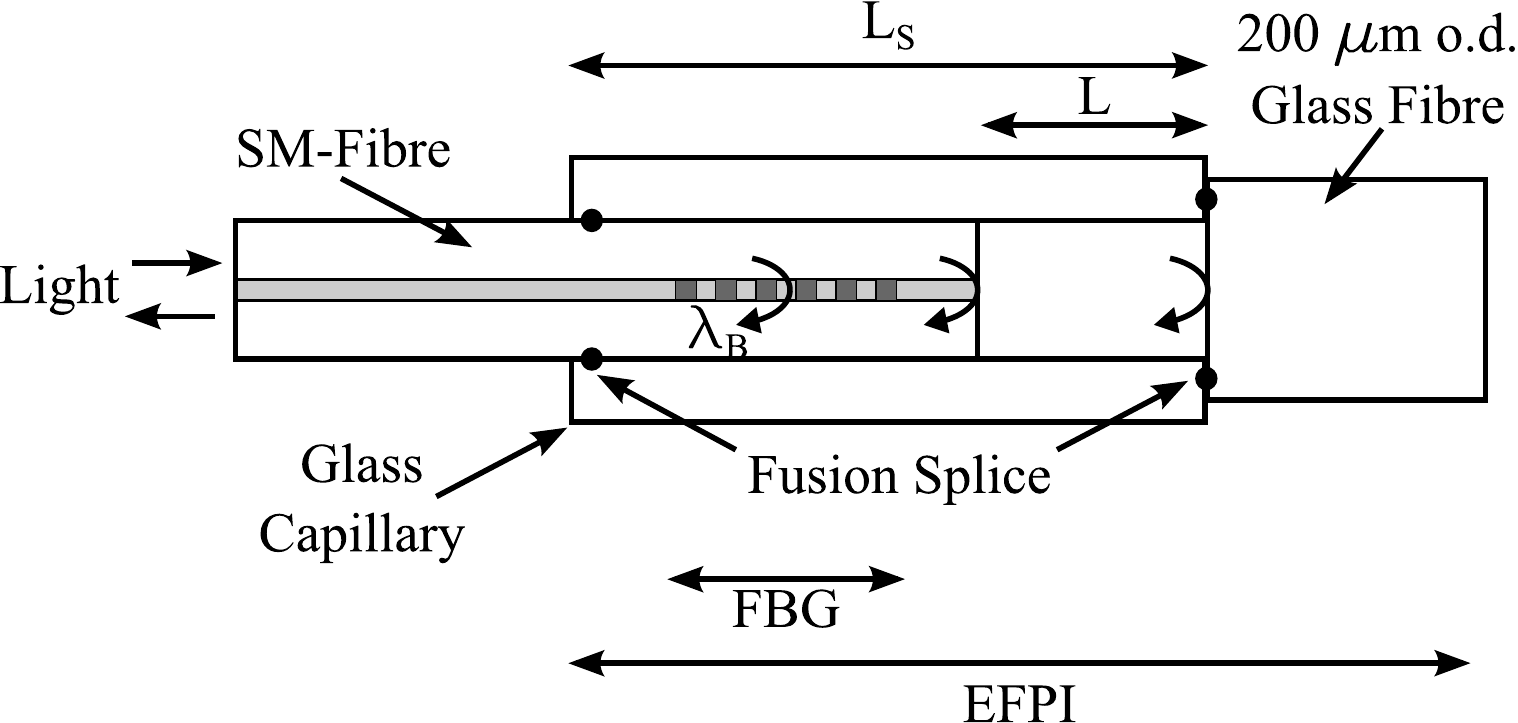}%
\caption{Sketch of the fibre optic sensor used for high pressure and temperature applications. The sensor consists of a combination of an Extrisic Fabry Perot Cavity (EFPI) and a Fibre Bragg Grating (FBG) for pressure and temperature sensing. $L$ denotes the length of the cavity, and $L_{S}$ the distance between the fusion splices.}%
\label{fig:sensor}%
\end{figure}

The fibre optic sensor (FOPS) used for this application consists of a miniature all-silica extrinsic Fabry-Perot cavity (EFPI) pressure sensor with an encapsulated Fibre Bragg Grating (FBG) for temperature sensing \cite{Bremer2010}. The sensor head is made of silica glass components by splicing a Single Mode (SM)-FBG and a 200~$\mu$m silica glass fibre to a silica glass capillary (Figure \ref{fig:sensor}). In addition, the 200~$\mu$m fibre was cleaved and polished using raw polishing paper several hundred micrometers from the glass capillary/200~$\mu$m fibre splice, in order to avoid light reflections at the outer surface of the 200~$\mu$m fibre. The length of the sensor is about 0.5~cm and its outer diameter is about 230~$\mu$m. Hence, the fibre optic sensor provides a simple, miniature and robust sensor configuration to measure pressure and temperature within a rock specimen.

The theoretical discussion of the sensing principle is based on \citet{Bremer2010}.

Incident light propagating to the head is reflected at the FBG for a wavelength equal to the Bragg wavelength $\lambda_{B}$ \cite{Kersey1997}:
\begin{equation}
\lambda_{B}=2n_{eff}\cdot\Lambda
\label{eq:fbg}
\end{equation}
with $n_{eff}$ being the refractive index of the fibre core material and $\Lambda$ the period of the grating. Wavelengths different than $\lambda_{B}$ propagate through the fibre and are reflected two times. Light becomes partly reflected at the entrance into the EFPI cavity (glass/air interface of the SM fibre). The transmitted light becomes reflected at the termination of the cavity (air/glass interface of the 200~$\mu$m fibre). Light, reflected from the end of the cavity is partly transmitted back into the SM fibre and interferes with light reflected from the first reflection at the entrance into the EFPI. The reflection coefficients of the glass/air and air/glass interface are low. Therefore, the function of the light interference can be calculated as \cite{Francis2008}:
\begin{equation}
I_{R}=I_{0}\cdot 2R(1+\cos(\varphi_{C}))
\label{eq:efpi}
\end{equation}
where $I_0$ and $I_R$ are the light intensities of the source and the reflected signal, respectively, and $R$ is the reflection coefficient of the glass/air and air/glass interface. $\varphi_{C;\lambda}$ is the phase shift between the reflected signals of wavelength $\lambda$. It is defined as:
\begin{equation}
\varphi_{C,\lambda}=\frac{4\pi n L}{\lambda}
\label{eq:phic}
\end{equation}
where $n$ is the refractive index of the EFPI cavity, $\lambda$ is the free space optical wavelength and $L$ is the EFPI cavity length. From this, the phase shift $\Delta \varphi_{C}$ between two wavelengths $\lambda_1$ and $\lambda_2$ can be calculated as:
\begin{equation}
\Delta \varphi_{C} =\varphi_{C,\lambda_{2}}-\varphi_{C,\lambda_{1}}= 4\pi n L\left(\frac{1}{\lambda_{2}}-\frac{1}{\lambda_{1}}\right)=4\pi n L \frac{\lambda_{1}-\lambda_{2}}{\lambda_{1} \lambda_{2}}
\label{eq:delphic}
\end{equation}
For a centre wavelength $\lambda_{1}$ and an adjacent wavelength $\lambda_{2}$ of equal phase, i.e.\ $\Delta \varphi_{C}=2 \pi$, the free spectral range $\lambda_{FSR,1}=\lambda_{1}-\lambda_{2}$ can be determined as:
\begin{flalign}
2\pi=\Delta \varphi_{C} & =4\pi n L \frac{\lambda_{1}-(\lambda_{1}-\Delta\lambda_{FSR,1})}{\lambda_{1} (\lambda_{1}-\Delta\lambda_{FSR,1})} \nonumber\\
\Delta\lambda_{FSR,1} & =\frac{\lambda_{1}^{2}}{2 n L + \lambda_{1}}
\label{eq:la2}
\end{flalign}

The glass capillary deforms under the influence of pressure. Hence, the length of the cavity changes, changing the interference pattern of the reflections. The cavity length change $\Delta L_{p}$ with respect to applied pressure $\Delta p$ can be expressed as \cite{Xu2005}:
\begin{equation}
\Delta L_{p}=\frac{L_{s}r_{0}^{2}}{E(r_{0}^{2}-r_{i}^{2})}(1-2\mu)\Delta p=a_{21}\Delta p
\label{eq:lp}
\end{equation}
where $\mu$ and $E$ are the Poisson's ratio and Young's modulus of the glass capillary, respectively. $L_{S}$ is the effective length of the pressure sensor, $r_{o}$ and $r_{i}$ are the inner and outer radius of the glass capillary. Besides the pressure sensitivity, the EFPI cavity is also sensitive to temperature, due to the thermal expansion of all glass components. The change of the cavity length as a result of temperature can be calculated as \cite{Xu2005}:
\begin{equation}
\Delta L_{T}=\left[(\beta_{C}-\beta_{F})L_{S}+\beta_{F}L+\frac{p}{T}a_{21}\right]\Delta T=a_{22}\Delta T
\label{eq:lt}
\end{equation}
where $\beta_{C}$ and $\beta_{F}$ are the coefficients of thermal expansion (CTE) of the glass capillary and the SM fibre, respectively. $p$ and $T$ are the pressure and temperature during sealing the EFPI cavity.

From previous experiments \cite{Bremer2010}, it is known that the pressure induced Bragg wavelength change is sufficiently low, when the FBG is entirely encapsulated in the glass capillary. Therefore, the pressure sensitivity of the FBG can be neglected and the coefficient $a_{11}$ in Equation \ref{eq:fbg-p} equals zero.
\begin{equation}
\Delta\lambda_{B,p}=a_{11}\Delta p
\label{eq:fbg-p}
\end{equation}

The temperature sensitivity of the FBG is based on an induced refractive index change and the CTE of the SM fibre. The shift of the Bragg wavelength $\lambda_{B}$ with respect to temperature can be expressed as \cite{Kersey1997}:
\begin{equation}
\Delta\lambda_{B,T}=\lambda_{B}\left(\beta_{F}+\frac{1}{n_{eff}}\frac{dn_{eff}}{dT}\right)\Delta T=a_{12}\Delta T,
\label{eq:lbt}
\end{equation}
where $dn_{eff}/dT$ is the thermo optic coefficient.

For the existing pressure and temperature relations (Equations \ref{eq:lp} to \ref{eq:lbt}) the following matrix representation can be
constructed:
\begin{equation}
\left[\begin{array}[m]{c}
	\Delta\lambda_{B}\\
	\Delta L
\end{array}\right]
=
\left[\begin{array}[m]{cc}
	a_{11}&a_{12}\\
	a_{21}&a_{22}
\end{array}\right]
\left[\begin{array}[m]{c}
	\Delta p\\
	\Delta T
\end{array}\right]
\label{eq:mat}
\end{equation}
where $a_{11}$ can be neglected, as stated above. 
\subsection{Sensor Interrogation System}
\begin{figure}%
\centering
\subfigure[Interrogation System.]{\label{fig:inter}\includegraphics[width=0.55\textwidth, angle=0]{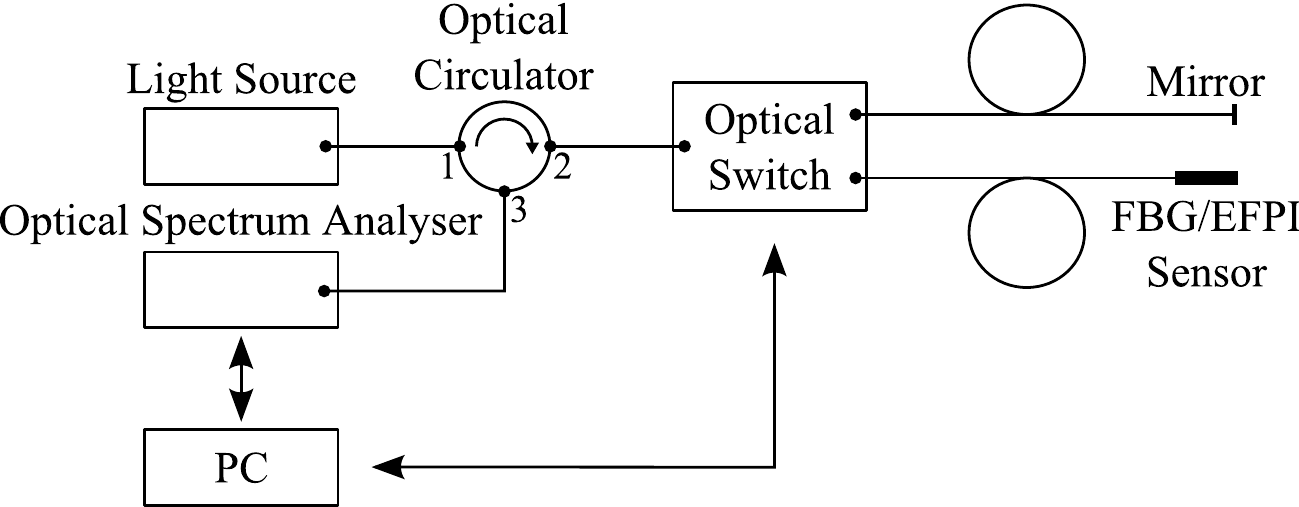}}\hspace{0.5 cm}
\subfigure[Normalized Spectrum.]{\label{fig:spec}\includegraphics[width=0.4\textwidth, angle=0]{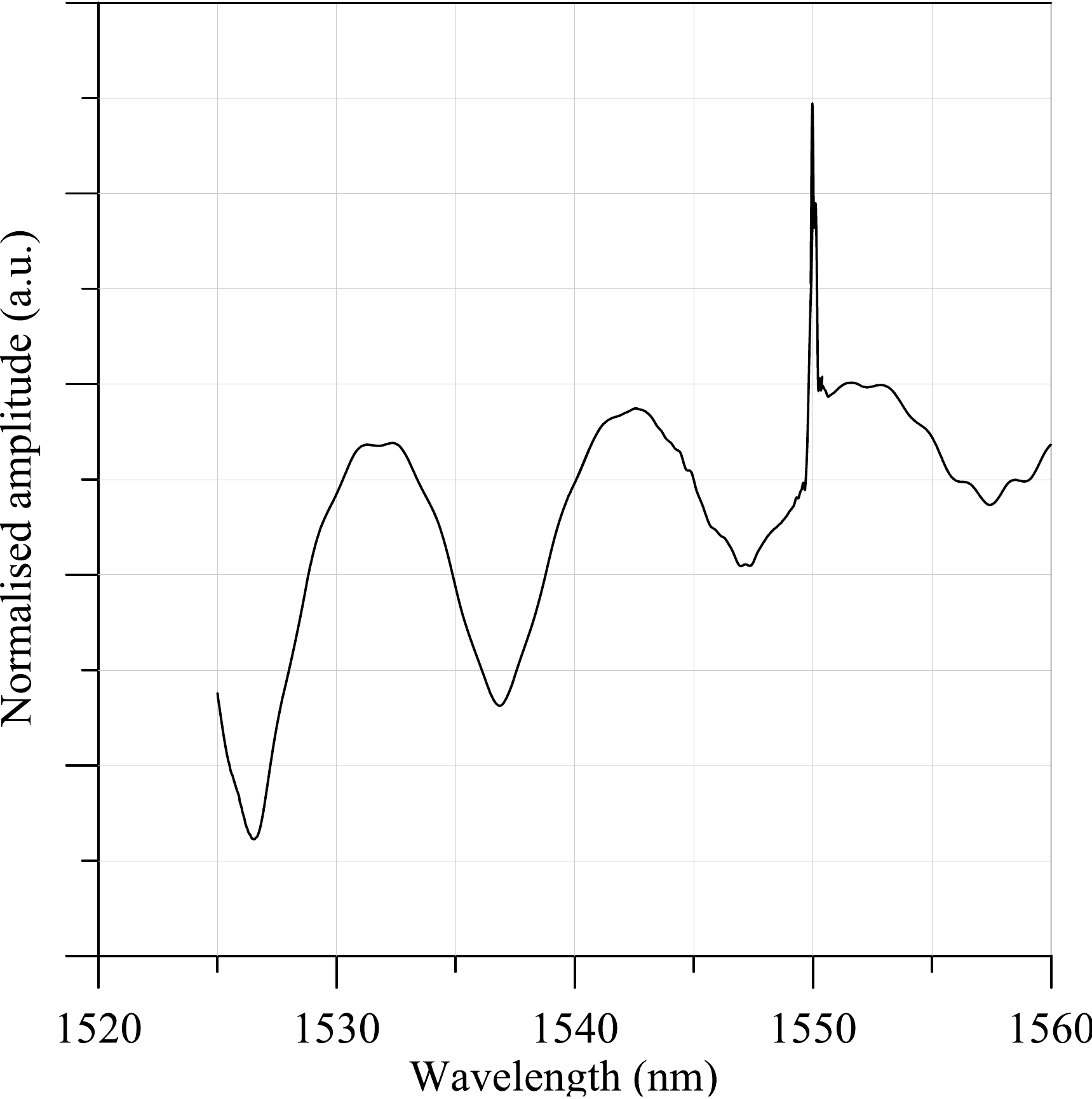}}

\caption{Sketch of the interrogation system (right) together with an exemplary optical spectrum (left) recorded with the novel p/T sensor.}%
\end{figure}

A schematic of the interrogation system is shown in Figure \ref{fig:inter}. The interrogation system consists of a broad-band light source (BBS) (INO FBS-C), an optical circulator, a fibre optic switch (JDS Fitel) and an optical spectrum analyser (OSA) (ANDO AQ6330).
Light from the BBS is guided through the optical circulator to the optical switch, which is used to interrogate the fibre optic sensor and an optical mirror, sequentially. The mirror signal is used for normalisation of the measured signal from the FOPS. The reflected light from the FOPS/mirror is then guided back to the optical circulator, from where the signal is transferred to the OSA. The OSA captures and normalises the reflected FOPS spectrum. A computer is used to acquire and analyse the spectrum and operating the switch. An example of a normalized spectrum is depicted in Figure \ref{fig:spec}.

As illustrated in Equations \ref{eq:lp} and \ref{eq:lt}, the EFPI cavity length changes with applied pressure and temperature, respectively. Consequently, the phase of the EFPI cavity changes with the measurands (Equation \ref{eq:phic}). The change in phase has been determined by the means of a Fast Fourier Transformation (FFT) algorithm and linearly fitted to the applied pressure. The FFT algorithm worked well for phase changes less or equal to $\Delta\varphi_{C}\leq 2\pi$. Phase ambiguities, however, occurred for phase changes larger than $2\pi$. Furthermore, a $2\pi$ periodic error has been imposed on the data. In order to correct this periodic error, calibration data for the entire pressure range has been linearly fitted. After subtracting the linear trend from the measured pressure, a sinusoidal function has been fitted to the data. This correction function was used to correct the measured data for the periodic error.

To unwrap the phase information, the change in the free spectral range $\Delta\lambda_{FSR,p}$ has been used. As the free spectral range was difficult to measure, an average free spectral range $\Delta\lambda_{FSR,p}^{avg}$ has been defined as the change in wavelength of the EFPI signal over the recorded wavelength range (1505-1585~nm). To calculate the change in $\Delta\lambda_{FSR,p}^{avg}$, a fractional fourier transformation (FRFT) \cite{Ozaktas2000} has been applied.

Within this study, only isothermal tests have been performed. Therefore, no temperature compensation has been applied.

\subsection{Implementation the FOPS into hp/hT triaxial cell}
\begin{figure}%
\centering
\subfigure[MTS Test Assembly.]{\label{fig:mts-set}\includegraphics[width=0.52\textwidth, angle=0]{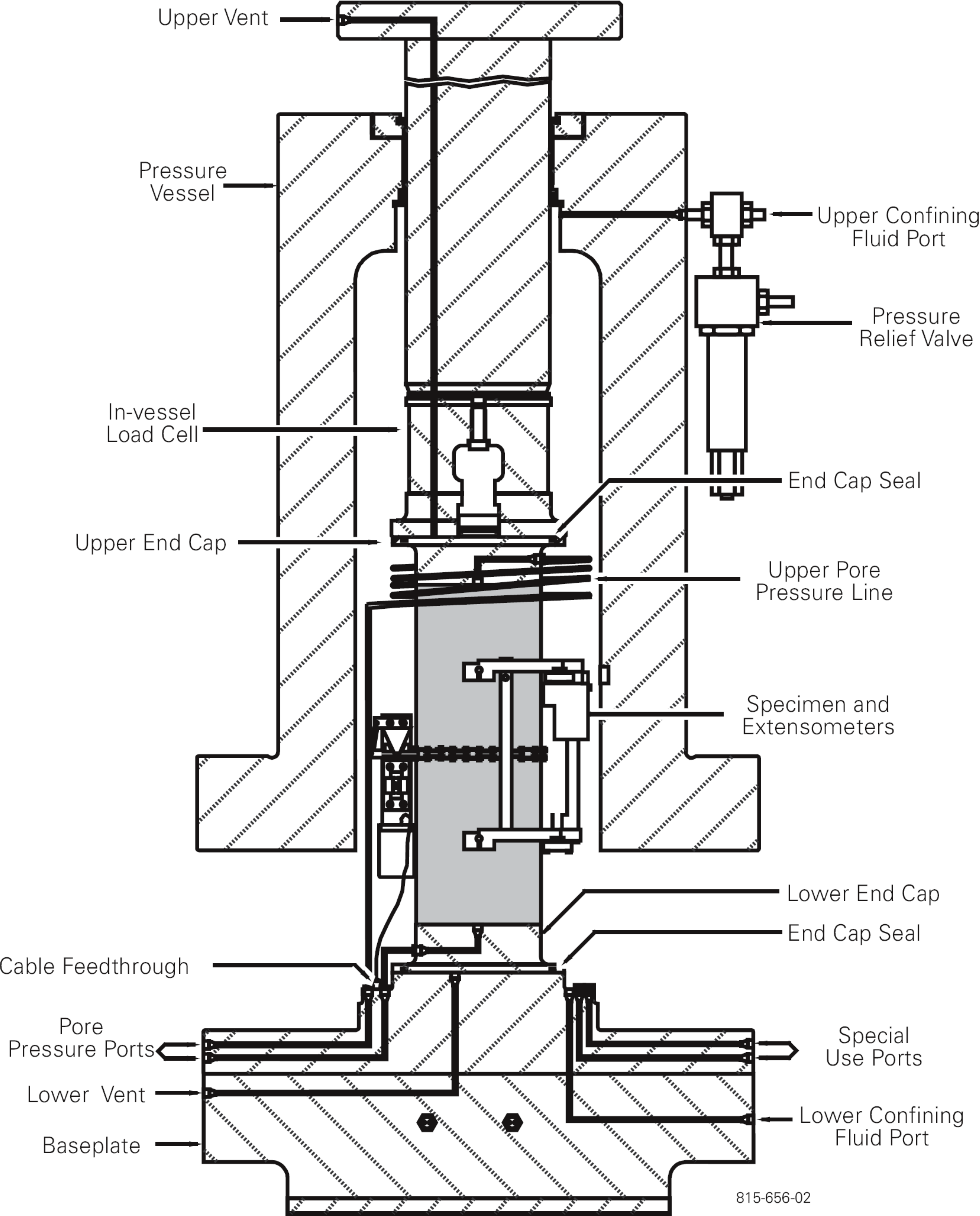}}\hspace{0.5 cm}
\subfigure[Sample within MTS.]{\label{fig:poro}\includegraphics[width=0.43\textwidth, angle=0]{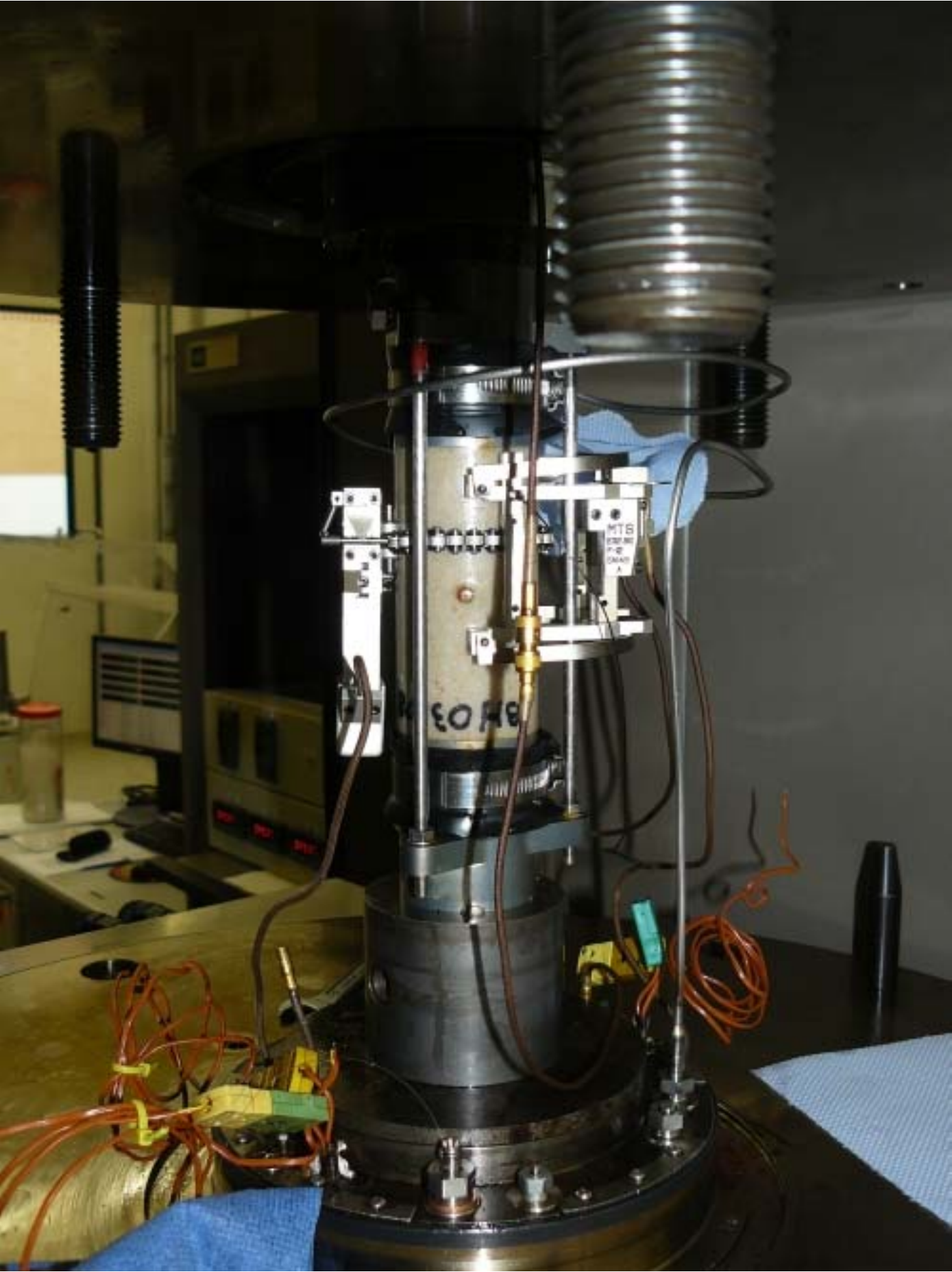}}

\caption{Laboratory test configuration showing a rock sample sample in the pressure chamber \cite{MTS2004} (left) and a sample mounted for porosity measurements (here, Bentheimer sandstone, right).}%
\label{fig:mts}
\end{figure}

In order to measure a pore pressure build-up within a rock specimen inside the pressurized chamber of a high pressure / high temperature test assembly, an optical fibre has to be fed into the high pressure chamber. Within this study a rock mechanical testing system, manufactured by MTS System Cooperation (Figure \ref{fig:mts}) was used. This test system allows for hydrostatic pressures up to 140~MPa and temperatures up to 200~$^{\circ}$C. Pressure and temperature within the chamber can be monitored using an external pressure sensor and three internal thermocouples (Type K), respectively (Figure \ref{fig:skem}). The thermocouples are located at the bottom, half the total length and the top of the sample within the pressure chamber. In order to implement the fibre optic sensor into the test assembly, a feed through for the optical fibre has been engineered. The lead fibre is fed through the base plate of the test assembly into the pressure chamber. Therefore, it was glued into a 1/8 inch stainless steel capillary tube using epoxy resin. This tube was mounted on the feed through and sealed using commercially available Swagelok fittings. Within the pressure chamber, the fibre optic sensor has been spliced to the lead fibre. This allows for an easy installation and replacement of sensors.

\section{Experimental Procedure}
In order to investigate the poroelastic behaviour geothermal reservoir rocks, a cylindrical sample of Flechtinger sandstone was prepared for the laboratory experiments. The Flechtinger sandstone has been chosen as an outcropping equivalent of the reservoir rock of Groß Schönebeck. It is a Lower Permian (Rotliegend) sedimentary rock quarried from an outcrop near Flechtingen, Germany. This Rotliegend sample is an arcosic litharenite containing varying amounts of quartz (55–65$\%$), feldspars (15–20$\%$), and rock fragments, mainly of volcanic origin (20–25$\%$). In addition, smaller amounts of clays are present, predominantly illite and chlorite \cite{Milsch2008}. The diameter of the sample was 5~cm and the was length 10~cm. Using Archimedes' principle, the initial porosity has been determined to be $\phi^i = 0.125$. For similar samples of the same rock block, initial permeability values $k^i$ between 0.3 and 1.3 mD have been measured. To allow for the sensor to be embedded within the sample, a hole with a diameter of 1.8~mm and a depth of approx. 20~mm has been drilled normal to the cylindrical surface. This hole is located at 4~cm distance from the top end face of the cylinder (see Figure \ref{fig:skem} and \ref{fig:skem-s}). The drill hole is a weak point and can potentially influence the results of the tests particularly in the case of deviatoric compression. Therefore, just hydrostatic compression was applied and the sensor itself was glued inside a 1/16 inch stainless steel capillary tube using epoxy resin. This capillary can resist the stress acting on its outer surface and provides an accurate pore pressure measurement at its interior which is connected to the pore space. Prior to the experiments, the sample was oven dried and afterwards saturated within a vacuum desiccator. By use of the dry and wet sample mass and the density of the saturant (distilled water), the initial pore volume $V_p^i$, the initial bulk volume $V_b^i$ and the initial porosity $\phi^i$ were calculated.

For this sample, three different isothermal hydrostatic compression tests have been performed to obtain different rock physical parameters. These are a drained hydrostatic compression test, an undrained hydrostatic compression test and an unjacketed hydrostatic compression test under in situ conditions. Prior to the drained compression test, the sample has been preconditioned. Table \ref{tab:tests} lists the different experiments performed . Three different set-ups have been used to measure the desired parameters.

\begin{table}%
\caption{List of different experiments. FEP denotes a shrinking tube made of fluorinated ethylene propylene.}
\label{tab:tests}
\begin{tabular}{lllll}
\toprule
Test&Measured&Derived&Sensor&Jacket\\
Type&Quantity&Quantity&      & Type\\
\midrule
Pretest&$V_p^i, V_b^i$&$\phi^i$&-&-\\
Preconditioning&&&Test&FEP\\
Drained&$\epsilon{}_v, dV_p$&$C_b, V_p, V_b, \phi, \alpha$&Calibration&FEP\\
Undrained&$p_p$&$B$&Pore Pressure&Neoprene\\
Unjacketed&$\epsilon{}_v$&$C_s, d\phi, \alpha, B$&Validation&-\\
\bottomrule
\end{tabular}
\end{table}	
\subsection{Sensor Test - Preconditioning}
Applying pressure to a rock sample, the sample is subject to reversible as well as irreversible deformations. To reduce the influence of an irreversible deformation as well as to minimize non-linear effects during the measurement, e.g.\ creep of the sample, it has to be preconditioned. A confining pressure cycle from 0.2 to 70~MPa with a period of 25~min has been applied, according to the procedure outlined by Hart and Wang \cite{Hart1995}. This procedure was repeated three times.

In order to simulate hydrostatic in situ conditions, the chamber was filled with hydraulic oil. To keep the sample from being contaminated, it was encapsulated in a shrinking tube made of fluorinated ethylene propylene (FEP jacket). The hydraulic system was connected to the pore system of the sample.

To minimize the influence of the drill hole for the sensor on the subsequent porosity measurement and to reduce the risk of a leakage when applying hydrostatic pressure, the hole has been plugged with a nail (Figure \ref{fig:skem-s}, filling the void space. To measure axial and lateral strain of the sandstone sample during compression, two axial extensometers and one radial chain extensometer were installed \cite{MTS2004} and the sample was mounted inside the pressure chamber (Figure \ref{fig:mts}). As the sample, the sensor has been placed within the pressure chamber and data has been recorded with a temporal resolution of 15~s. Using data from the preconditioning stage, the interrogation system and the sensor response was tested (Figure \ref{fig:precon}).

\subsection{Sensor Calibration - Drained Compression Test}
The drained compression test has been performed directly after the preconditioning phase on the jacketed rock specimen. As confining pressure $p_c$ increases, and pore pressure $p_p$ is held constant, the change in pore volume $dV_p$ and specimen volumetric strain $\epsilon_v=\epsilon_a+2\epsilon_c$ were measured. Only this test is required to calculate the Biot coefficient $\alpha$ and porosity $\phi$ by the direct method. The Biot Coefficient $\alpha$ can be defined as the ratio of the change in pore volume $dV_p$ and the unit change of bulk volume $dV_b$ under drained conditions \cite{Kuempel1991}. As the MTS system provides axial and circumferential strain measurements only, the change in bulk volume has been replaced by the volumetric strain $dV_b=V_b^id\epsilon_v$.
\begin{equation}
\alpha=\left(\frac{dV_p}{dV_b}\right)_{p_p}=\left(\frac{dV_p}{V_b^id\epsilon_v}\right)_{p_p}
\end{equation}

For the porosity measurement a distinction between the definition of Lagrange's porosity $\phi_L$ and of Euler's porosity $\phi_E$, see Equation \ref{eq:porosities}, has to be made. Lagrange's porosity refers to the initial configuration, i.e. the ratio of the pore volume defined in the deformed state and the bulk volume in the initial state. In contrast, Euler's porosity refers to the deformed configuration, i.e. the ratio of the pore volume and the bulk volume, both in the deformed state \cite{Mainguy2002}. In the following, we are referring to the Euler's porosity, only.

\begin{equation}
\phi_L=\frac{V_p}{V_b^i}
\phi_E=\frac{V_p}{V_b},
\label{eq:porosities}
\end{equation}

There are four different compressibilities defined for porous rock, which relate changes in either pore volume $V_p$ or bulk volume $V_b$ to changes in either confining pressure $p_c$ or pore pressure $p_p$ \cite{Zimmerman1986}. During the drained hydrostatic compression test, the pore pressure inside the sample is kept constant. Therefore, the variations of the bulk volume and pore volume with applied confining pressure yield the drained bulk compressibility $C_b$ and the drained pore compressibility $C_p$, respectively \cite{Ghabezloo2009}.

\begin{equation}
C_b=\frac{1}{K_b}=-\frac{1}{V^i_b}\left(\frac{\partial V_b}{\partial p_e}\right)_{p_p},C_p=\frac{1}{K_p}=-\frac{1}{V^i_p}\left(\frac{\partial V_p}{\partial p_e}\right)_{p_p}
\label{eq:comp1}
\end{equation}

During the drained compression test, the sensor has been calibrated. Therefore, the sensor was kept within the pressure chamber and a pressure cycle from 0.2 to 70~MPa has been applied. Confining pressure has been increased and subsequently reduced stepwise in 5~MPa steps, except for the low pressure range, where steps at 0.5, 1.5 and 2.5~MPa have been performed. At all pressure steps, 10 optical spectra from the FOPS have been recorded. Using the pressure readings from the pressure chamber, the sensor was calibrated (see Figure \ref{fig:calib_valid}).

\subsection{Sensor Pore Pressure Measurement - Undrained Compression Test}\label{subsubsec:skem}
For fully saturated rocks under undrained conditions, the pore pressure $p_p$ change is a function of the applied mean stress. Under hydrostatic conditions the mean stress is equal to the confining pressure $p_c$. The ratio of pore pressure change to the applied confining pressure change defines the isotropic Skempton coefficient $B$ (\cite{Brown1975};\cite{Green1986}). The pore pressure change is related to confining pressure change by the following equation \cite{Skempton1954}:

\begin{equation}
B=\frac{dp_p}{dp_c}.
\end{equation}

\begin{figure}%
\centering
\subfigure[Sample Preparation.]{\label{fig:skem}\includegraphics[width=0.39\textwidth, angle=0]{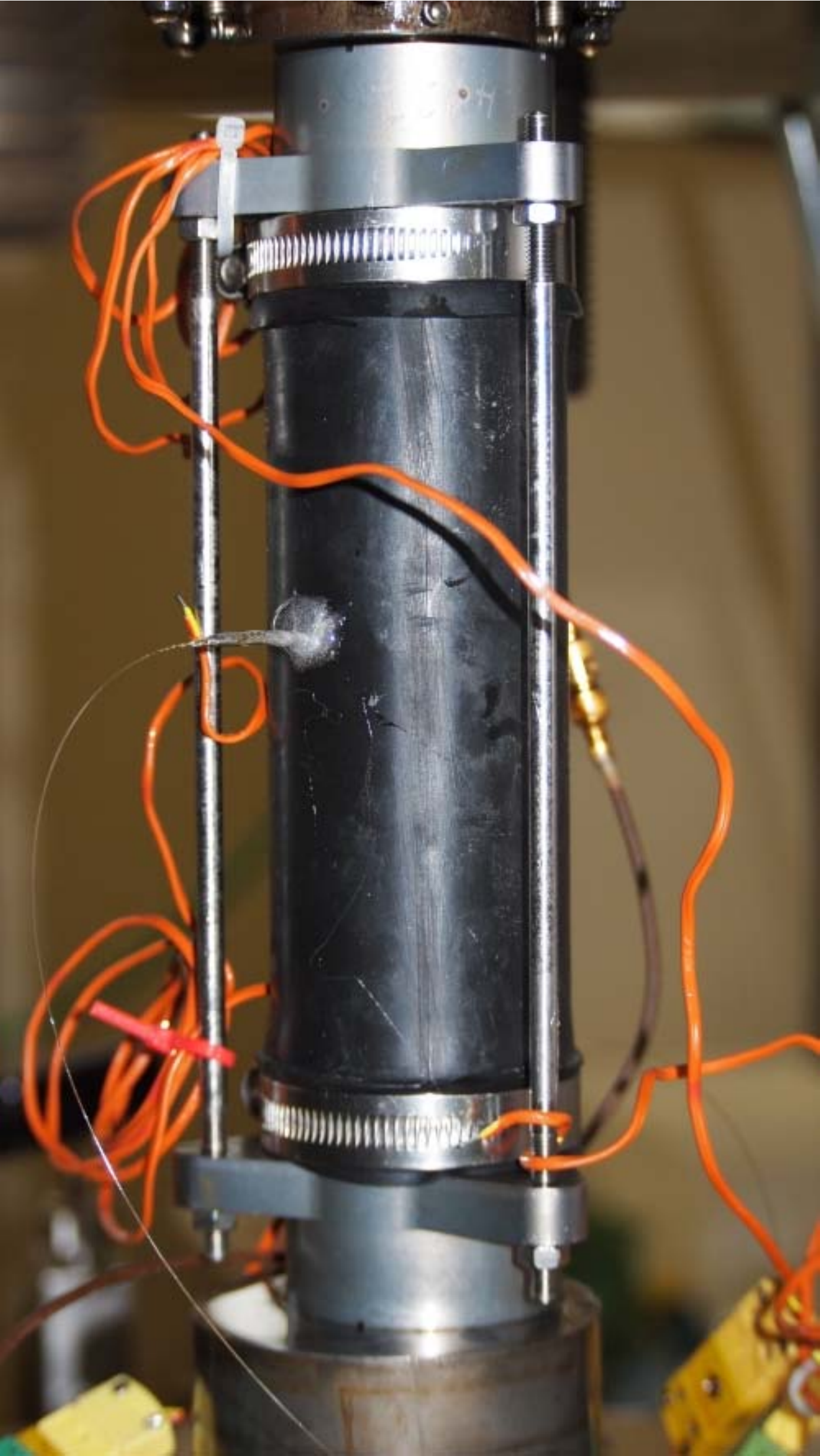}}\hspace{0.5 cm}
\subfigure[Sketch.]{\label{fig:skem-s}\includegraphics[width=0.56\textwidth, angle=0]{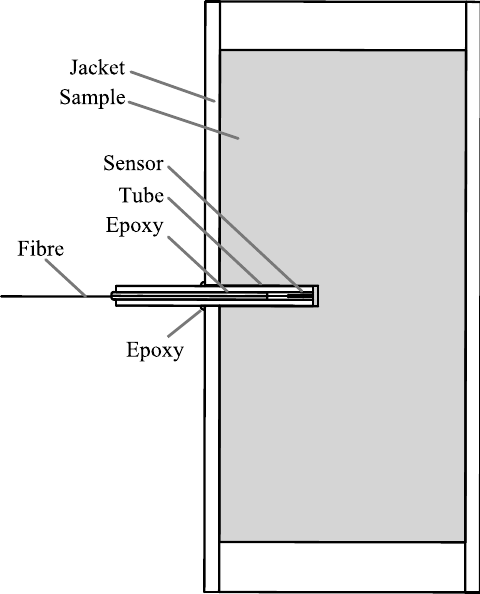}}

\caption{Picture showing the prepared sample for the pore pressure measurements (left) together with a cross section of the embedded sensor (right). See Section \ref{subsubsec:skem} for further details.}%
\end{figure}

For pressure measurements within the rock specimen, the fibre optic sensor has been embedded in the rock specimen (Figure \ref{fig:skem}). The hydraulic system was disconnected, the extensometers and the nail were removed together with the FEP jacket and replaced by a Neoprene jacket and two blind end caps on either side of the cylindrical sample. Using a needle, a hole has been poked through the jacket at the position of the drill-hole. The sensor has been embedded through this hole approx 2~cm deep into the rock sample, leaving a void space of 12~mm$^{3}$ which is less than 0.05$\%$ of the pore volume. The resin inside the steel capillary seals the pore pressure from the confining pressure and it reduces the void space within the tube.

Applying pressure to the sample, the flexible Neoprene jacked seals the feed-through of the optical sensor. To reduce the probability of leakage at the feed-through, a flexible epoxy resin has been disposed close to the feed-through.

For the pore pressure determination, confining pressures from 0.2 to 70~MPa have been applied with a rate of 10~MPa/h. Again, pressure readings from the sensor have been acquired with a temporal resolution of 15~s.

\subsection{Sensor Data Validation - Unjacketed Compression Test}

In order to validate the sensor calibration and for measuring the unjacketed bulk compressibility $C_s$, an unjacketed hydrostatic compression test was performed. During the unjacketed compression test, the increments in confining pressure and pore pressure are equal and applied simultaneously. Therefore, the effective pressure remains constant $p_e=p_c-p_p=0$ at this conditions. The variations in bulk volume and pore volume with respect to the applied pressure are defined as unjacketed bulk compressibility $C_s$ and unjacketed pore compressibility $C_\phi$ \cite{Zimmerman1986}.
\begin{equation}
C_s=\frac{1}{K_s}=-\frac{1}{V^i_b}\left(\frac{\partial V_b}{\partial p_p}\right)_{p_e},C_\phi=\frac{1}{K_\phi}=-\frac{1}{V^i_p}\left(\frac{\partial V_p}{\partial p_p}\right)_{p_e}
\label{comp2}
\end{equation}

The experimental evaluation of the unjacketed pore compressibility is difficult, since fluid compressibility and the compressibility of external tubing and the pore pressure generator has to been taken into account \cite{Ghabezloo2009}. In order to measure the unjacketed bulk compressibility, the Neoprene jacket has been removed and the two axial extensometers and the radial chain extensometer have been attached to the sample.

By means of the unjacketed bulk compressibility in conjunction with the measured drained bulk compressibility during the drained compression test, the Biot coefficient can be determined by an indirect method (\cite{Biot1957},\cite{Nur1971}):
\begin{equation}
\alpha=1-\frac{C_s}{C_b}
\end{equation}

Furthermore, the unjacketed bulk compressibility can be used to calculate changes in porosity due to changes in effective pressure \cite{Jaeger2007}. This relation arises from a theory of hydrostatic poroelastics developed by Carroll and Katsube \cite{Carroll1983}.
\begin{equation}
d\phi_E=-\left[(1-\phi^i)C_b-C_s\right]d(p_e)
\label{eq:dphi_jaeger}
\end{equation}

In Equation \ref{eq:dphi_jaeger} the assumption was made, that the unjacketed bulk compressibility is identical with the unjacketed pore compressibility and the compressibility of the solid matrix, $C_s=C_\phi=C_m$. This assumption is valid for homogeneous and isotropic media at the micro-scale. Without this simplification an additional term regarding the unjacketed pore compressibility due to a pore pressure change must be taken into account \cite{Ghabezloo2009}.

\begin{equation}
d\phi_E=-\left[(1-\phi^i)C_b-C_s\right]d(p_e)+\phi^i\left(C_s-C_\phi\right)dp_p
\label{eq:dphi_ghabezloo}
\end{equation}

By use of the direct porosity measurement during the drained compression test and the porosity data determined by the indirect method (Equation \ref{eq:dphi_jaeger}), the additional term $\phi^i\left(C_s-C_\phi\right)dp_p$ can be calculated. In particular, by use of the porosity data from the drained compression test in conjunction with the unjacketed compression test, the unjacketed pore compressibility $C_\phi$ can be estimated.

Finally, the unjacketed bulk compressibility can be used for calculating the Skempton coefficient $B$. \citet{Mesri1976} derived an equation which relates the Skempton coefficient to $C_f, C_b$ and $C_s$, where the assumption was made that the solid matrix compressibility is identical with the unjacketed bulk compressibility $C_m=C_s$. In \cite{Jaeger2007}, a similar equation can be found replacing the unjacketed bulk compressibility $C_s$ by the compressibility of the solid matrix $C_m$.
\begin{equation}
B=\frac{\frac{K_f}{\phi^i}\left(1-\frac{K_b}{K_s}\right)}{\frac{K_f}{\phi^i}\left(1-\frac{K_b}{K_s}\right)
+K_b\left(1-\frac{K_f}{K_s}\right)}=\frac{C_b-C_s}{\phi^i(C_f-C_s)+(C_b-C_s)}
\end{equation}

During the unjacketed compression test, the fibre optic sensor has been removed from the drill hole and placed within the pressure chamber. The set-up is therefore similar to the calibration experiment and sensor pressure readings can be validated using the information about the confining pressure. Two pressure cycles from 0.2 to 70~MPa have been applied and data has been recorded with a temporal resolution of 15~s. Data from this experiment has been used to validate pressure readings from the second experimental step.

\section{Results}
\label{sec:res}
As mentioned in the introduction, only data, relevant for the fibre optic sensor, will be presented. The rock physical implication of the results will be discussed in a separate publication.

\subsection{Sensor Test - Preconditioning}
\begin{figure}%
\includegraphics[width=1\textwidth]{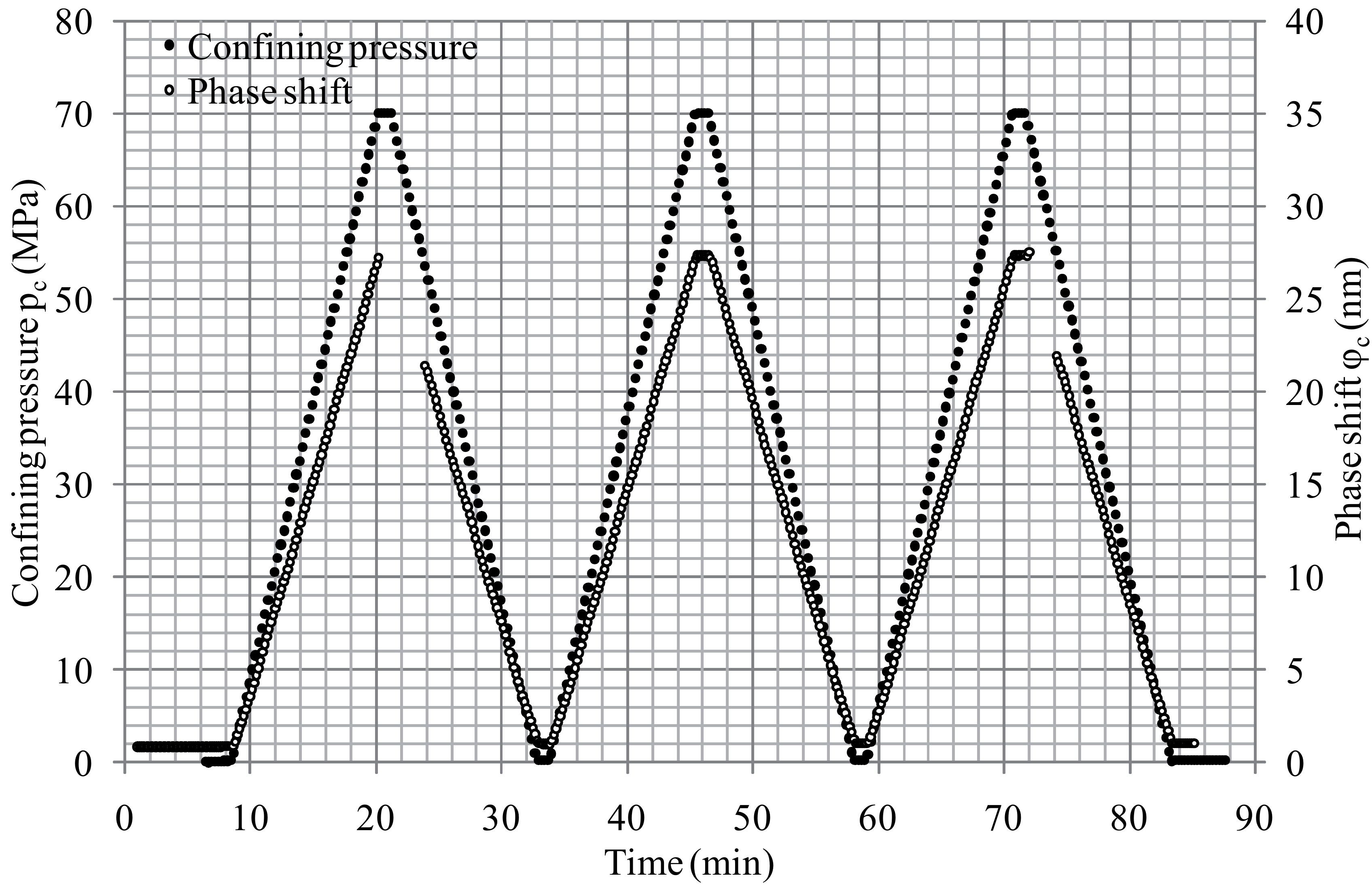}%
\caption{Confining pressure $p_c$ and phase shift $\varphi_c$ of the interference signal vs. time. The missing data of down ramp one and three corresponds with a reset of the optical spectrum analyser device.}%
\label{fig:precon}%
\end{figure}

During the preconditioning phase, the sensor has been tested on functionality and the pressure response has been evaluated. Figure \ref{fig:precon} shows the pressure ramps over time that have been applied to the sample. The measured phase shift has been plotted as well. The data shows a good correlation of the measured phase shift with the applied pressure. The temporal response was below the time resolution of the data acquisition.

\subsection{Sensor Calibration - Drained Compression Test}
\begin{figure}%
\centering
\subfigure[Calibration and Validation.]{\label{fig:calib_valid}\includegraphics[width=0.475\textwidth, angle=0]{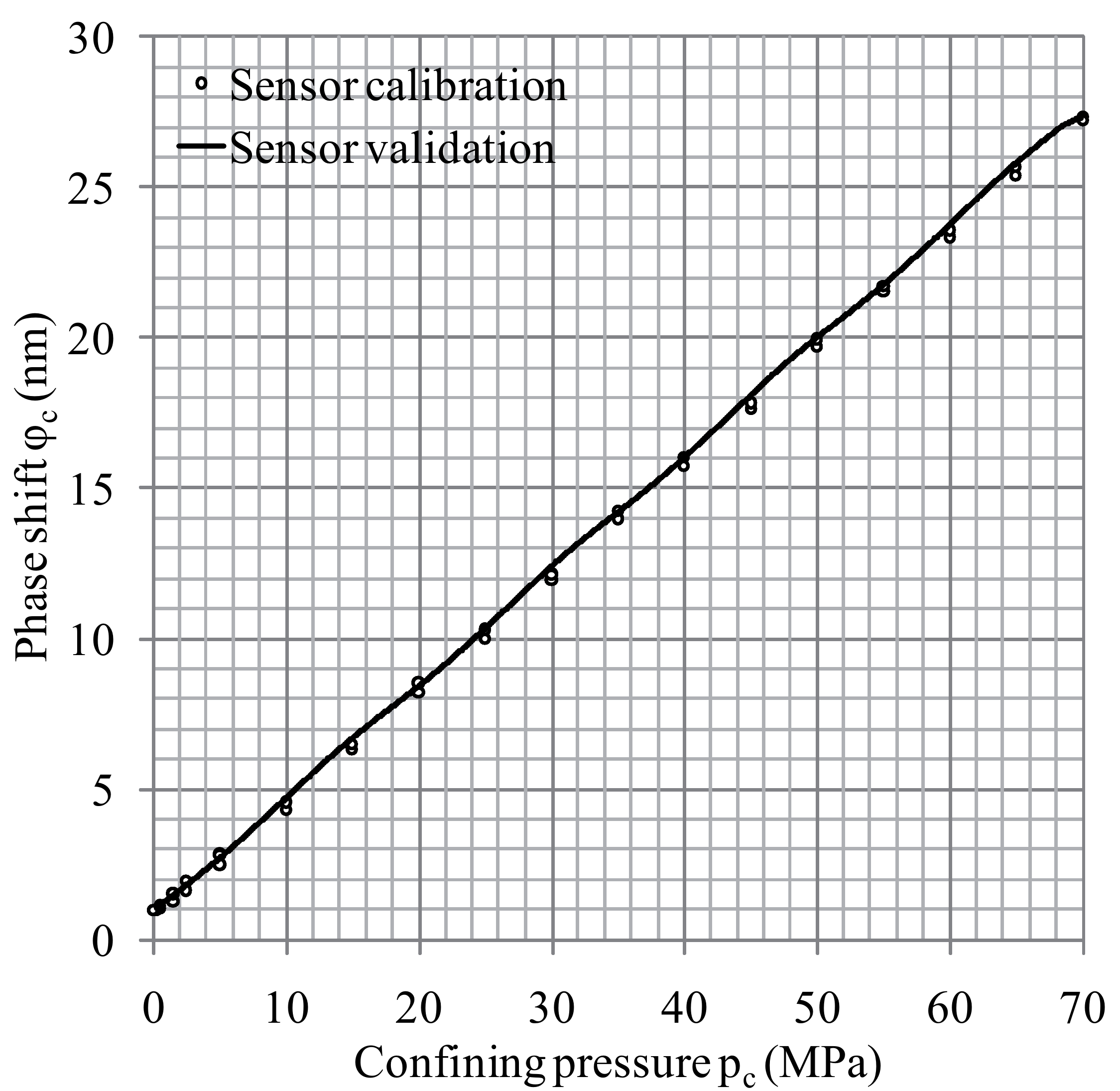}}\vspace{0.5 cm}
\subfigure[Measurement.]{\label{fig:meas}\includegraphics[width=0.475\textwidth, angle=0]{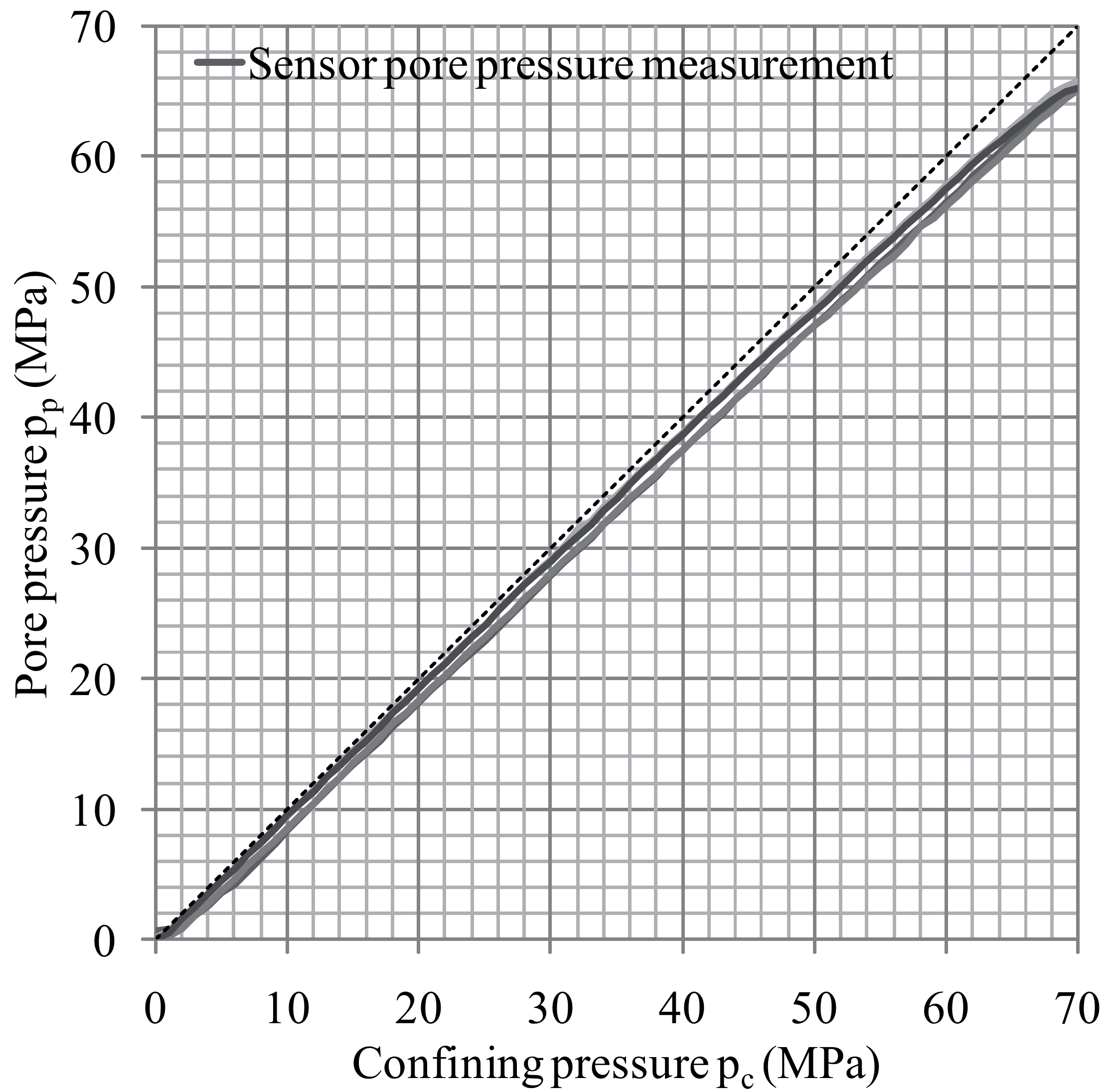}}
\caption{Left: Phase shift $\varphi_c$ vs. confining pressure $p_c$ of the sensor calibration during the drained compression test (white circles) in comparison with the sensor validation during the unjacketed compression test (continuous line). Data has been fitted linearly for calibrating the sensor. Right: Measured pore pressure vs. applied confining pressure during the undrained compression test. The pore pressure is displayed with solid lines. Data has been corrected by a sinusoidal correction function. For comparison, the dashed line shows data for the case that the measured pore pressure is equal to the applied confining pressure. }%
\end{figure}

Figure \ref{fig:calib_valid} shows the measured phase shift at different confining pressure values. For this experimental set-up, an absolute accuracy of $\pm~0.75$~MPa ($\approx 1\%$ full scale) has been calculated for the applied pressure range from 0.2 to 70~MPa by linear fitting.

\subsection{Sensor Pore Pressure Measurement - Undrained Compression Test}
The measured pore pressure in relation to the applied confining pressure has been measured for the entire range from 0.2 to 70~MPa. Based on the primary data processing a periodic error for the pore pressure data could be analysed. To mitigate this error a sinus correction function was determined and applied. As shown in Figure \ref{fig:meas} a maximum effective pressure $p_e$ of 5~MPa has been measured at 70~MPa confining pressure.

\subsection{Sensor Validation - Unjacketed Compression Test}
Data from the unjacketed compressibility test has been used to validate the measured pressure data. As seen in Figure \ref{fig:calib_valid}, the difference in pressure determination prior and after the in situ pore pressure measurement is less than 0.1~MPa.

\section{Discussion and Conclusions}
\label{sec:disc}
For the first time, pore pressure information was acquired directly within the pore space of a rock specimen by means of a fibre optic sensor. The sensing system proved to be successful and a valuable tool to measure in situ pressure conditions during a hydrostatic compression test.

Several different tests have been performed to measure the most relevant poro-elastic parameters of a rock specimen. The tests have been arranged, so that all relevant steps of sensor test, calibration, measurement and data validation were integrated. By this combination of tests it is possible to determine the Biot coefficient $\alpha$, the Skempton coefficient $B$ and the porosity $\phi$ by direct and indirect method. Furthermore, drained bulk compressibility $C_b$, drained pore compressibility $C_p$, unjacketed bulk compressibility $C_s$ and unjacketed pore compressibility $C_\phi$ can be measured.

Data from the preconditioning experiment has been used to evaluate the sensor performance. The sensor proved to be able to operate in the desired pressure range. Furthermore, a good temporal response has been determined over the entire pressure range from 0.2 to 70~MPa. The linear fit of the calibration data resulted in an absolute accuracy of 0.75~MPa for the applied experimental set-up. This could be improved by sinusoidal fitting of the resulting calibration curve. 

The pore pressure $p_p$ data, recorded during the jacketed hydrostatic compression test, proved to be sufficiently accurate for the determination of the Skempton coefficient $B$. Finally, data validation during the unjacketed hydrostatic compression test ensured, that measured pressures where correct within the measurement accuracy. Using the sinusoidal fitting, the reproducibility between both measurements was determined to be better 0.1 MPa.

\section{Summary}
\label{sec:sum}
To understand the behaviour of rocks under changing load or temperature conditions, the determination of physical parameters like pore pressure or temperature within the pore space is essential. Within this study, the implementation of a novel fibre optic point sensor for pressure and temperature determination into a high pressure / high temperature triaxial cell is presented. The sensor used within this study consists of a miniature all-silica fibre optic Extrinsic Fabry-Perot Interferometer (EFPI) sensor which has an embedded Fibre Bragg Grating (FBG) reference sensor element to determine temperature and pressure directly at the point of measurement. For the first time, pressure was measured directly within the pore space of a Flechtinger sandstone specimen during a hydrostatic compression test at up to 70~MPa.

\section{Outlook}
\label{sec:out}
Within this study, in situ pore pressure data was acquired at a single point of measurement within the rock specimen under investigation. For future work, however, several sensors along the sample can be used to measure effective pressure conditions under varying load conditions. Data can be used to measure e.g.\ fracture permeabilities, fracture propagation or the internal stress state of a specimen.

For non-isothermal measurements, the steel capillary tube has to be exchanged by a ceramic capillary with a thermal conductivity similar to the bulk thermal conductivity of the rock. For a capillary tube with a different thermal conductivity, the temperature information form the inside of the sample would be biased by the outside conditions.

In order to increase the accuracy of the sensor, the sensor dimensions will be optimized for the desired pressure range and an optimized data processing algorithm will be used.

\section*{Acknowledgements}
This work has been performed in the framework of the GeoEn-Phase~2 project and funded by the Federal Ministery of Education and Research [BMBF, 03G0767A], the IRCSET (Embark Initiative) and Science Foundation Ireland [SFI/ENEF662]. Furthermore, the authors are grateful to C. Schmidt-Hattenberger for lending the necessary fibre optic equipment, L. Liebeskind for the support in performing the experiments as well as J. Schr\"{o}tter and R. Giese for the help in modifying the MTS set-up. For the fruitful discussion regarding the theory of poroelasticity, the authors would like to thank S. Ghabezloo and A. Hassanzadegan.

\printnomenclature

\bibliographystyle{model1-num-names}
\bibliography{ijrmms}

\begin{thebibliography}{23}
\expandafter\ifx\csname natexlab\endcsname\relax\def\natexlab#1{#1}\fi
\providecommand{\bibinfo}[2]{#2}
\ifx\xfnm\relax \def\xfnm[#1]{\unskip,\space#1}\fi
\bibitem[{Green and Wang(1986)}]{Green1986}
\bibinfo{author}{H.~G. Green}, \bibinfo{author}{H.~F. Wang},
\newblock \bibinfo{title}{Fluid pressure response to undrained compression in
  saturated sedimentary rock.},
\newblock \bibinfo{journal}{Geophysics} \bibinfo{volume}{51, 4, 948 - 956.}
  (\bibinfo{year}{1986}).
\bibitem[{Ghabezloo and Sulem(2010)}]{Ghabezloo2010}
\bibinfo{author}{S.~Ghabezloo}, \bibinfo{author}{J.~Sulem},
\newblock \bibinfo{title}{Effect of the volume of the drainage system on the
  measurement of undrained thermo-poro-elastic parameters},
\newblock \bibinfo{journal}{International Journal of Rock Mechanics and Mining
  Sciences} \bibinfo{volume}{47} (\bibinfo{year}{2010}) \bibinfo{pages}{60 --
  68}.
\bibitem[{Wissa(1969)}]{Wissa1969}
\bibinfo{author}{A.~E.~Z. Wissa},
\newblock \bibinfo{title}{Pore pressure measurement in satureted stiff soils.},
\newblock \bibinfo{journal}{J. Soil. Mech.} \bibinfo{volume}{95}
  (\bibinfo{year}{1969}) \bibinfo{pages}{1063 -- 1073}.
\bibitem[{Schmidt-Hattenberger et~al.(2003)Schmidt-Hattenberger, Naumann, and
  Borm}]{Schmidt-Hattenberger2003}
\bibinfo{author}{C.~Schmidt-Hattenberger}, \bibinfo{author}{M.~Naumann},
  \bibinfo{author}{G.~Borm},
\newblock \bibinfo{title}{Fiber bragg grating strain measurements in comparison
  with additional techniques for rock mechanical testing},
\newblock \bibinfo{journal}{IEEE Sensors Journal} \bibinfo{volume}{3}
  (\bibinfo{year}{2003}) \bibinfo{pages}{50 -- 55}.
\bibitem[{Bremer et~al.(2010)Bremer, Lewis, Leen, Lochmann, Mueller, Reinsch,
  and Schroetter}]{Bremer2010}
\bibinfo{author}{K.~Bremer}, \bibinfo{author}{E.~Lewis},
  \bibinfo{author}{B.~Leen, G.and~Moss}, \bibinfo{author}{S.~Lochmann},
  \bibinfo{author}{I.~Mueller}, \bibinfo{author}{T.~Reinsch},
  \bibinfo{author}{J.~Schroetter},
\newblock \bibinfo{title}{Fibre optic pressure and temperature sensor for
  geothermal wells},
\newblock in: \bibinfo{booktitle}{Sensors, 2010 IEEE}, \bibinfo{address}{Opt.
  Fibre Sensor Res. Centre, Univ. of Limerick, Limerick, Ireland}, p.
  \bibinfo{pages}{538}.
\bibitem[{Kersey et~al.(1997)Kersey, Davis, Patrick, LeBlanc, Koo, Askins,
  Putnam, and Friebele}]{Kersey1997}
\bibinfo{author}{A.~Kersey}, \bibinfo{author}{M.~Davis},
  \bibinfo{author}{H.~Patrick}, \bibinfo{author}{M.~LeBlanc},
  \bibinfo{author}{K.~Koo}, \bibinfo{author}{C.~Askins},
  \bibinfo{author}{M.~Putnam}, \bibinfo{author}{E.~Friebele},
\newblock \bibinfo{title}{Fiber grating sensors},
\newblock \bibinfo{journal}{Journal of Lightwave Technology}
  \bibinfo{volume}{15} (\bibinfo{year}{1997}) \bibinfo{pages}{1442 -- 1463}.
\bibitem[{Yu et~al.(2008)Yu, Yin, and Ruffin}]{Francis2008}
\bibinfo{editor}{F.~T. Yu}, \bibinfo{editor}{S.~Yin}, \bibinfo{editor}{P.~B.
  Ruffin} (Eds.), \bibinfo{title}{Fiber Optic Sensors},
  \bibinfo{publisher}{Taylor \& Francis Group}, \bibinfo{edition}{2} edition,
  \bibinfo{year}{2008}.
\bibitem[{{Xu} et~al.(2005){Xu}, {Wang}, {Cooper}, {Pickrell}, and
  {Wang}}]{Xu2005}
\bibinfo{author}{J.~{Xu}}, \bibinfo{author}{X.~{Wang}}, \bibinfo{author}{K.~L.
  {Cooper}}, \bibinfo{author}{G.~R. {Pickrell}}, \bibinfo{author}{A.~{Wang}},
\newblock \bibinfo{title}{{Miniature fiber optic pressure and temperature
  sensors}},
\newblock in: \bibinfo{editor}{{M.~A.~Marcus, B.~Culshaw, \& J.~P.~Dakin}}
  (Ed.), \bibinfo{booktitle}{Society of Photo-Optical Instrumentation Engineers
  (SPIE) Conference Series}, volume \bibinfo{volume}{6004} of
  \textit{\bibinfo{series}{Presented at the Society of Photo-Optical
  Instrumentation Engineers (SPIE) Conference}}, pp. \bibinfo{pages}{7--12}.
\bibitem[{Ozaktas et~al.(2000)Ozaktas, Zalevsky, and Kutay}]{Ozaktas2000}
\bibinfo{author}{H.~M. Ozaktas}, \bibinfo{author}{Z.~Zalevsky},
  \bibinfo{author}{M.~A. Kutay}, \bibinfo{title}{The Fractional Fourier
  Transform, with Applications in Optics and Signal Processing},
  \bibinfo{publisher}{Wiley}, \bibinfo{year}{2000}.
\bibitem[{{MTS Systems Cooperation}(2004)}]{MTS2004}
\bibinfo{author}{{MTS Systems Cooperation}}, \bibinfo{title}{Rock Mechanics and
  Concrete Mechanics Testing System - Technical Description},
  \bibinfo{type}{Technical Report}, MTS Systems Cooperation,
  \bibinfo{year}{2004}.
\bibitem[{Milsch et~al.(2008)Milsch, Blöcher, and Engelmann}]{Milsch2008}
\bibinfo{author}{H.~Milsch}, \bibinfo{author}{G.~Blöcher},
  \bibinfo{author}{S.~Engelmann},
\newblock \bibinfo{title}{The relationship between hydraulic and electrical
  transport properties in sandstones: An experimental evaluation of several
  scaling models},
\newblock \bibinfo{journal}{Earth and Planetary Science Letters}
  \bibinfo{volume}{275} (\bibinfo{year}{2008}) \bibinfo{pages}{355 -- 363}.
\bibitem[{Hart and Wang(1995)}]{Hart1995}
\bibinfo{author}{D.~J. Hart}, \bibinfo{author}{H.~F. Wang},
\newblock \bibinfo{title}{Laboratory measurements of a complete set of
  poroelastic moduli for berea sandstone and indiana limestone},
\newblock \bibinfo{journal}{Journal of Geophysical Research, B, Solid Earth and
  Planets} \bibinfo{volume}{100} (\bibinfo{year}{1995}) \bibinfo{pages}{17741
  -- 17751}.
\bibitem[{K\"{u}mpel(1991)}]{Kuempel1991}
\bibinfo{author}{H.-J. K\"{u}mpel},
\newblock \bibinfo{title}{Poroelasticity: parameters reviewed},
\newblock \bibinfo{journal}{Geophysical Journal International}
  \bibinfo{volume}{105} (\bibinfo{year}{1991}) \bibinfo{pages}{783--799}.
\bibitem[{Mainguy and Longuemare(2002)}]{Mainguy2002}
\bibinfo{author}{M.~Mainguy}, \bibinfo{author}{P.~Longuemare},
\newblock \bibinfo{title}{Formulation du couplage partiel entre un simulateur
  r\'eservoir et un simulateur de g\'eom\'ecanique},
\newblock \bibinfo{journal}{Oil \& Gas Science and Technology - Rev. IFP}
  \bibinfo{volume}{57} (\bibinfo{year}{2002}) \bibinfo{pages}{355 -- 367}.
\bibitem[{Zimmerman et~al.(1986)Zimmerman, Somerton, and King}]{Zimmerman1986}
\bibinfo{author}{R.~W. Zimmerman}, \bibinfo{author}{W.~H. Somerton},
  \bibinfo{author}{M.~S. King},
\newblock \bibinfo{title}{Compressibility of porous rocks},
\newblock \bibinfo{journal}{Journal Of Geophysical Research}
  \bibinfo{volume}{VOL. 91, NO. B12} (\bibinfo{year}{1986}) \bibinfo{pages}{765
  -- 777}.
\bibitem[{Ghabezloo et~al.(2009)Ghabezloo, Sulem, and
  Saint-Marc}]{Ghabezloo2009}
\bibinfo{author}{S.~Ghabezloo}, \bibinfo{author}{J.~Sulem},
  \bibinfo{author}{J.~Saint-Marc},
\newblock \bibinfo{title}{Evaluation of a permeability-porosity relationship in
  a low-permeability creeping material using a single transient test},
\newblock \bibinfo{journal}{International Journal of Rock Mechanics and Mining
  Sciences} \bibinfo{volume}{46} (\bibinfo{year}{2009}) \bibinfo{pages}{761 --
  768}.
\bibitem[{Brown and Korringa(1975)}]{Brown1975}
\bibinfo{author}{R.~J.~S. Brown}, \bibinfo{author}{J.~Korringa},
\newblock \bibinfo{title}{On the dependence of the elastic properties of a
  porous rock on the compressibility of the pore fluid.},
\newblock \bibinfo{journal}{Geophysics} \bibinfo{volume}{40, 608-–616.}
  (\bibinfo{year}{1975}).
\bibitem[{Skempton(1954)}]{Skempton1954}
\bibinfo{author}{A.~W. Skempton},
\newblock \bibinfo{title}{The pore pressure coefficient a and b.},
\newblock \bibinfo{journal}{G\'{e}otechnique} \bibinfo{volume}{4, 143–-147.}
  (\bibinfo{year}{1954}).
\bibitem[{Biot and Willis(1957)}]{Biot1957}
\bibinfo{author}{M.~A. Biot}, \bibinfo{author}{D.~G. Willis},
\newblock \bibinfo{title}{The elastic coefficients of the theory of
  consolidation},
\newblock \bibinfo{journal}{Journal of Applied Mechanics} \bibinfo{volume}{24}
  (\bibinfo{year}{1957}) \bibinfo{pages}{594 -- 601}.
\bibitem[{Nur and Byerlee(1971)}]{Nur1971}
\bibinfo{author}{A.~Nur}, \bibinfo{author}{J.~D. Byerlee},
\newblock \bibinfo{title}{An exact effective stress law for elastic deformation
  of rock with fluids},
\newblock \bibinfo{journal}{Journal Of Geophysical Research}
  \bibinfo{volume}{VOL. 76, NO. 26} (\bibinfo{year}{1971}) \bibinfo{pages}{6414
  -- 6419}.
\bibitem[{Jaeger et~al.(2007)Jaeger, Cook, and Zimmerman}]{Jaeger2007}
\bibinfo{author}{J.~C. Jaeger}, \bibinfo{author}{N.~G.~W. Cook},
  \bibinfo{author}{R.~W. Zimmerman}, \bibinfo{title}{Fundamentals of Rock
  Mechanics}, \bibinfo{publisher}{Blackwell Publishing Ltd.},
  \bibinfo{edition}{4th edition} edition, \bibinfo{year}{2007}.
\bibitem[{Carroll and Katsube(1983)}]{Carroll1983}
\bibinfo{author}{M.~M. Carroll}, \bibinfo{author}{N.~Katsube},
\newblock \bibinfo{title}{The role of terzaghi effective stress in linearly
  elastic deformation},
\newblock \bibinfo{journal}{Journal of Energy Resources Technology}
  \bibinfo{volume}{105} (\bibinfo{year}{1983}) \bibinfo{pages}{509 -- 511}.
\bibitem[{Mesri et~al.(1976)Mesri, Adachi, and Ullrich}]{Mesri1976}
\bibinfo{author}{G.~Mesri}, \bibinfo{author}{K.~Adachi}, \bibinfo{author}{C.~R.
  Ullrich},
\newblock \bibinfo{title}{Pore pressure response in rock to undrained change in
  all-round stress.},
\newblock \bibinfo{journal}{G\'{e}otechnique} \bibinfo{volume}{25, 317- 330.}
  (\bibinfo{year}{1976}).

\end{thebibliography}

\end{document}